\documentclass{sigchi}

\CopyrightYear{2023}
\setcopyright{acmcopyright}
\conferenceinfo{Towards an Inclusive and Accessible Metaverse at ACM CHI'23,}{April  25--30, 2023, Hamburg, Germany}
\usepackage{balance}       % to better equalize the last page
\usepackage{graphics}      % for EPS, load graphicx instead 
\usepackage[T1]{fontenc}   % for umlauts and other diaeresis
\usepackage{txfonts}
\usepackage{mathptmx}
\usepackage{hyperref}
\hypersetup{pdflang={en-US}}
\usepackage{color}
\usepackage{booktabs}
\usepackage{textcomp}

\usepackage{microtype}        % Improved Tracking and Kerning
\usepackage{ccicons}          % Cite your images correctly!
\usepackage{todonotes}

\def\plaintitle{Disconnected from Reality: Do the core concepts of the metaverse exclude disabled individuals?%Crashing through the snow: Has science-fiction held back the metaverse?
}

\def\emptyauthor{}
\def\plainkeywords{Authors' choice; of terms; separated; by
  semicolons; include commas, within terms only; this section is required.}

\makeatletter
\def\url@leostyle{%
  \@ifundefined{selectfont}{
    \def\UrlFont{\sf}
  }{
    \def\UrlFont{\small\bf\ttfamily}
  }}
\makeatother
\def\pprw{8.5in}
\def\pprh{11in}

\definecolor{linkColor}{RGB}{6,125,233}
\hypersetup{%
  pdftitle={\plaintitle},
  pdfauthor={\emptyauthor},
  pdfkeywords={\plainkeywords},
  pdfdisplaydoctitle=true, % For Accessibility
  bookmarksnumbered,
  pdfstartview={FitH},
  colorlinks,
  citecolor=black,
  filecolor=black,
  linkcolor=black,
  urlcolor=linkColor,
  breaklinks=true,
  hypertexnames=false
}

\begin{document}

\title{\plaintitle}

\numberofauthors{1}
\author{%
  \alignauthor{Mark Quinlan\\
    \affaddr{Department of Computer Science, University College London}\\
    \affaddr{London, United Kingdom}\\
    \email{m.quinlan@ucl.ac.uk}}\\
}

\maketitle

\begin{figure*}
  \centering
  \includegraphics[width=2\columnwidth]{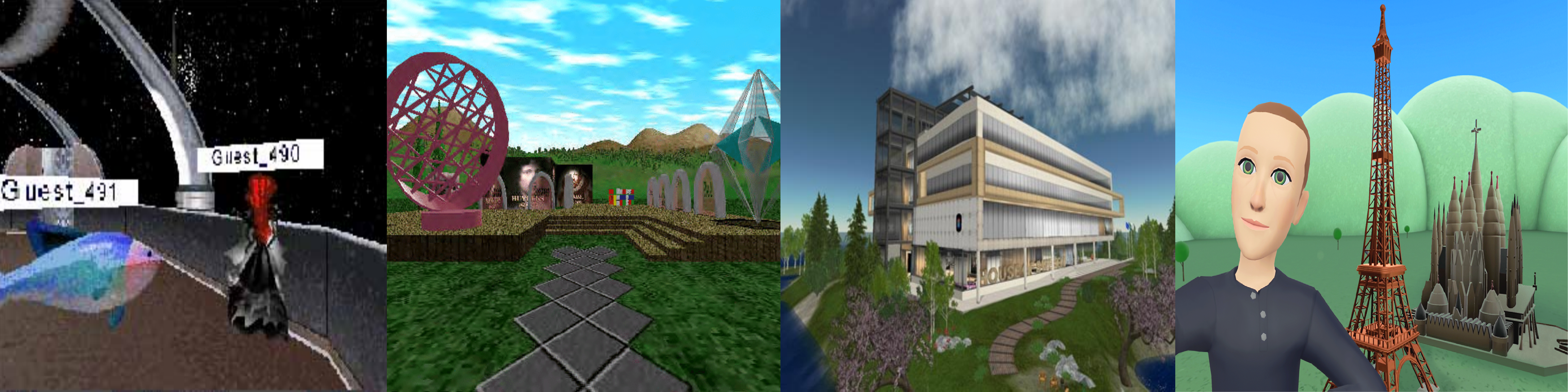}
  \caption{Examples of virtual worlds. From the far-left, we see a `fish' interact with an `alien' (author's impression). Both are user-controlled avatars in \textit{Worldschat}, circa 1994. Centre-left, we see a virtual art-show in \textit{Active Worlds}, circa 1995. Centre-right, we see the Digital Swedish Embassy in \textit{Second-Life}, circa 2007. Far-right, we see Meta CEO Mark Zuckerberg take a `selfie' within \textit{Horizon Worlds}, circa 2021. }~\label{fig:figure1}
\end{figure*}

\begin{abstract}
Commercially-driven metaverse development has been driven by philosophical and science fiction concepts. Through translating these concepts into products, the developers may have inadvertently excluded individuals with disabilities from this new expanded reality. This ideologically-driven development is presented in this paper through a brief background of what we see as the most influential of these concepts, and explain how these might affect disabled individuals wishing to engage with said products. It is our hope that these ideas prompt conversation on future inclusivity access from the concept stage of future metaverse development.
\end{abstract}

% ACM Classfication



% Author Keywords
% Print the classficiation codes

\section{Virtual Worlds: A (very) brief introduction}

Novel technologies are not derived from the ether; rather, they emerge from knowledge, linguistic and technological contexts \cite{chesher1994colonizing}. Virtual worlds (referred to in this paper in the plural as metaversae) are no exception. Briefly, a metaverse is a simulated environment, often populated by users who can interact with each other and their environment using avatars \cite{sparkes2021metaverse}. The scope and features of virtual worlds vary greatly, and in this paper we look to those on those that focus on interpersonal and economic relationships, such as Meta's Horizon Worlds. Consider the linguistics of the terminology at work; The virtual, or `not quite real' \cite{shields2005virtual} can be traced back in concept at least as far as Plato. In his cave allegory, Plato argued that if one had never set foot in the real world, but had only seen projections of it on the cave wall of the outside world, then that world would constitute reality \cite{plato}. A subjective reality, which in contrast to what we know is only virtual to the cave-dweller. In many ways the metaversae developed since the 2010's mirror this concept, not trying to mirror reality closely, but instead offering a configurable but impoverished version of it.

Moving beyond the ancient Greeks, the renaissance also had its opinions on virtual worlds. In 1641 Descartes published his seminal work \textit{Meditations on First Philosophy}, which contained a thought-experiment on virtual worlds. In the treatise, Descartes created a thought-experiment that must have been terrifying to consider for the first time; what if, instead of being in the `real' world, he was trapped in a world created entirely to deceive his senses? In this metaverse, a malevolent creator was responsible for creating these illusions, and they were so perfect that Descartes described that ``the earth, colors, figures, sound, and all other external things are naught". Perfect illusions from a creator which had perfect access to his senses. Descartes therefore thought-up the ultimate metaverse, one completely indistinguishable from the physical reality \cite{descartes}.

In the modern era, Deleuze ascribes more explicit meaning to the concept of the virtual. Deleuze replaces the concept of the possible with the virtual, stating that the virtual is already `real' and doesn't need to await realization. Genesis of thought is the same as actualization of the virtual, thus any given thought or concept is already realized in the mind \cite{pearson2005reality}. These concept we see closely in our chosen fictional examples, in the next section.

\subsection{From concept to fiction: inspiring the builders}

Science fiction, which was heavily inspired by philosophy \cite{bridle2022ways}, is a narrative genre where many novels rely on "realistic" and easy-to-follow narratives to describe virtual worlds. We provide two seminal examples. In the first, Gibson in his 1982 novel \textit{Neuromancer} creates a vision of `cyberspace'\footnote{Cyberspace has since has become synonymous with descriptions of global computer networks such as the Internet at every level.}, which was seen as `a new universe' --- parallel to the physical, but created by the digital~\cite{gibson1982neuromancer}. Gibson's visions of cyberspace allow for the navigate through it as if it were a physical domain, using tools that directly transferred as metaphors form the `real' to his vision of the metaverse \cite{betz2013analogical}. In Gibson's prose, cyberspace is accessed by “jacking in” to the user's brain via a VR interface; they experience a brightly colored 3-D landscape of data structures and virtual physical locations, described as “ice” and “data castles” that run with code. It is a metaverse with users and AIs.

In the second, we look towards Stephenson's work \textit{Snow Crash} \cite{stephenson2003snow}. Stephenson created the term Metaverse (capitalized in the book, as per his penchant for utilizing brand-names) to refer to a world of interconnected virtual reality systems, accessible anywhere. The visual elements are vivid and surreal, and users control avatars to explore and engage in various activities. The universe mirrors aspects of \textit{Neuromancer}, but unlike the “ice and data castles” of Gibson's cyberspace, the visual elements of the metaverse are far more vivid and concrete, and filled with surrealistic imagery.

The above examples are of course non-exhaustive of the genre, but were chosen for four reasons. First, both works' ideas bear similarities to the philosophical theories we described, in that we see the development of cyberspace within it from an impoverished form of reality that nonetheless entrances its protagonists, warping their sense of reality, through to the ideas of the metaverse transcending what Deleuze thought limited the `possible', with thought genesis often being enough to bring about change. 

Second, they have had tremendous influence on almost every area that has a part to play in the development of metaversae, ranging from their etymology (many academic definitions of virtual world, or virtual reality can be traced to these works \cite{lanier1992virtual, heim1993metaphysics}), through to governments \cite{benedikt1991cyberspace}, media \cite{betz2013analogical}, software developers \cite{kraus2022facebook}, investors and entrepreneurs\footnote{Many Silicon valley founders have spoken directly of their appreciation for the two novels. For example Google's co-founder Sergey Brin \cite{feloni}.} \cite{song2021explication}. And third, these provided an \textit{au fait} description of their metaversae, to the extent that you could argue that the two works provided the blueprints for the metaversae described below. And fourth, Stephenson's work highlights one narrative for disabled individuals, showing the power of the metaverse to transcend reality, as well as narrative tropes surrounding disability \footnote{We refer here to the sub-plot involving the character Ng.}. And yet broad representations of disability in science fiction do not necessarily translate into product development. 

\subsection{From fiction to fact: genesis metaverse}

Both today and during the 1990s when many early metaversae were released, developers predicated their strategies on a specific interpretation of metaversae, referring to the above works and acting as `a virtual mirror world to physical reality' \cite{golf2022embracing}. Figure \ref{fig:figure1} highlight four of these formative and influential metaversae, which are representative of the genre. The first, Worldscat, was released originally in 1994, only two years after \textit{Snow Crash} was published. WorldsChat offered a 3D graphical experience, and within it, users could create and customize their own environment, and interact with a community. To further encourage expression, WorldsChat provided tools for creating 3D objects and for manipulating elements of the environment \cite{damer1997inhabited}, some of which can not be seen in more modern interpretations of metaversae. Active Worlds (1995) had many of the same features, and included scripting functionality for timed events using a virtual time zone, paving the way for the `virtuality' espoused by Deleuze and shown in \textit{Snow Crash}.

Second Life introduced open-source functionality \cite{galanis2009open}, detailed visuals, and an in-world economy with the option of exchanging virtual goods for real currency. Users could create and sell 3D objects, craft stories, and monetize their efforts \cite{kaplan2009fairyland}. We can consider Second Life the first metaverse that gained mainstream attention, despite not developing a user-base of significant size, countries such as Sweden created the first `digital embassies' (see: Fig \ref{fig:figure1} and companies such as Stellantis (then: Fiat Auto) used it to host product launches \cite{lucatorto2009clil}.  Ball termed Second Life a `protometaverse' the template which most future metaversae developers would most likely follow \cite{ball2022metaverse}. 

Meta's (formerly known as Facebook) Horizon World's (far-right, Fig\ref{fig:figure1}) can be seen as the next iteration of Second Life, and is supported by a huge organization, whose products serve more than one-third of the world's population \cite{meta}. Meta is a proponent of using VR technology to further the concept of the metaverse, and is spending vast sums in support of this. These technologies could potentially lead to Descartes' perfect VR system and the sensation of `embodiment' the feeling of a virtual body existing in a digital world. And yet Meta's vision is sanitized in comparison to what came before in fiction and practice, lacking the freedom to create a character that fits with the variety that humanity offers\footnote{To the extent that even legs, or the lack thereof, became a topic of discussion \cite{metasucks}.}

\section{The needs of the many: Excluding individuals from virtual worlds}

\subsection{Physical exclusion}

It is in this section where we explicitly divide between metaversae and their access points. Metaversae in almost all extant and science-fiction cases\footnote{See \textit{Star Trek: The Next Generation's} Holodeck as an example of inclusivity within metaversae access.} rely on VR technology. As the vision for metaversae has expanded, so has investment into tools which may have no immediate facilities for disabled individuals. Corporations such as Google, Microsoft and HTC spend billions on developing mixed-reality tools which rely exclusively on vision and haptics, and Facebook re-branded itself to Meta as if to state explicitly it was following in Stephenson's vision\footnote{In a wide-ranging interview, Meta CEO Mark Zuckerberg describes his vision of the metaverse in a similar manner to what is contained in the novel, before avoiding a question regarding said similarity \cite{theverge2}. At the same time, Stephenson has disavowed himself of Meta's interpretation \cite{disavow}.} \cite{theverge}. Gerling \& Spiel levy the accusation that Virtual Reality (VR) tends to prioritize an ideal `corporeal standard' and neglects to properly attend to the needs of alternatives in its design. Facebook’s Oculus technology is an example of this standard, being designed around normative cognitive, spatial and kinetic capabilities \cite{spiel2021purpose}. There are some studies to corroborate this with practical facts; for example one by the Disability Project, who stated that the tools in use for many VR games using Oculus Technology did not even include height adjustments, reinforcing the notion of an ideal body type as a pre-requisite for use \cite{survey}. 

The tools that do get developed tend to follow reasonably explicitly the vision laid out by the above fiction examples, especially with regard to access\footnote{Both novels discuss tools that could be translated to products like Meta's Oculus headset. See: https://www.oculus.com}. Egliston et al \cite{egliston2021critical, golf2022embracing} sees this as an `unseen' harm, which is especially prescient in the case of disabled individuals. It thus becomes clear why such arbitrary restrictions originated, and how they contribute to a increasingly unequal and ableist environment. 

\subsection{Exclusion of identity}

Our world is built from physical, the other is built from philosophy and fiction. How do we express identity in between these worlds? To begin, in the physical world, identity begins through existence \cite{bukatman1993terminal}. In the metaverse, perhaps it begins through Descartes `embodiment'. This embodiment is not just the physical sensation, it is the belief in what you are experiencing \cite{park2022metaverse}. Physical exclusion therefore leads to identity exclusion, which can have severe psychological effects on an individual \cite{park2022metaverse}. Positive examples of normative individuals experiencing such identity development in metaversae can be found in Japan, where the `VTuber' phenomena has led to new identities being developed that are seen by its metaverse participants not as separate, but wholly as extensions of themselves which were inexpressible before their metaverse was developed \cite{bredikhina2020avatar, lu2021more}. We could argue that in this case we see digital avatars that its users understand as part of their `true' and immutable selves, and aid in the development of a social identity \cite{stets2003sociological}.

\subsection{Social Exclusion}

Post-Facebook acquired Oculus frames its technology as part of the everyday, something easily incorporated into daily life. This has caused it to distance itself from its once-common reputation of being a niche technology for enthusiasts and researchers \cite{egliston2022oculus}. This discourse fails to recognize the needs of those that do not conform to the `corporeal standard', leaving them even more invisible \cite{egliston2022oculus}. 

One example of a relevant social implication may lie in the way in which these technologies rely so heavily on data in order to facilitate access to metaversae. This leads us to the economic motive of tool development. For example, Facebook can be described as an advertising platform. Through tracking and monitoring users' social activity, it creates a large collection of data that can be used to facilitate automated transactions between advertisers and Facebook's ad network, enabling wealth creation through data \cite{egliston2021critical}. Thus Oculus reflects not only the concept that Meta took from the Stephenson's Metaverse, it reflects their economic reality. In response to this, recent research has shifted its focus to examine the data-intensive nature of VR technologies. These technologies encompass novel and advanced forms of algorithmic decision-making, for example work by Cho on eye-tracking and blink rates as a determinant for cognitive load \cite{cho2021rethinking}. Many of these data-intensive areas tend to lead towards what Egliston et al terms `datafied governance' \cite{egliston2021critical}. In these situations, at the social and institutional level data is the prime determinant for decision making with impacts ranging from fairness and bias in access provision through to social implications that prevent disabled users from being able to interact with wider society. All of which we can see in the way power is stratified in the metaversae of Gibson and Stephenson. As a practical example, we already see this in the use of AI language classifiers which more often classify texts using terms pertaining to disability identity as toxic and negative \cite{whittaker2019disability}. Thus even when not physically explicitly excluded, it may be that simply not being able to provide data (or data not fitting with normative ideals) will lead to social discrimination or exclusion, and exclude those with disabilities from a domain that promises to extend the very nature of reality for those meeting the corporeal ideal, leaving others trapped in a Cartesian VR-nightmare.  

\subsection{Economic Exclusion}

If modern metaversae have stopped offering open source tools to alter reality, such as seen in Active Worlds, they instead provide a vibrant economy, where individualism can be purchased, and those with the requisite skills can participate in this economy. Furthermore, we have seen the conceptual merging of of metaverse and web3 concepts lead to abstract economic activity that has yet to be fully explored. 

In what this paper terms the \textit{Metaverse Exclusive Economic Zone} (MEEZ), physical, identity and social exclusion naturally leads to a dearth of economic opportunities. These do not necessarily need to be abstract in origin. For example, work by Lyons looked at talent acquisition in the metaverse, and the transfer of work environments from the physical to the metaverse \cite{lyons2022talent}. if these become mainstream, it is not difficult to use how these may in turn deprive disabled individuals from being able participate in economic activities that would otherwise potentially still have been open to them. 

This paper believes that these areas are not mutually exclusive, they compound and reinforce issues within each other, as we so often see in the physical world \cite{walsh2003social}. Transferring these issues into metaversae, through denial of service will only lead to exacerbating them in both worlds. 

\section{Interactive storytelling as a potential avenue for inclusivity}

This paper believes that through the confluence of philosophy and science fiction the metaverse has been shaped into what it is today. In such a case, could perhaps an interdisciplinary solution sourced from these areas be the answer? Again, there has been prior work in the field that could qualify as a cyber-physical technological solution. Work by IBM to develop a spoken-word command-line version of second-life for legally-blind individuals highlights a prior ambitious attempts to provide access to metaversae that otherwise remain closed \cite{folmer2009textsl}, by incorporating new avenues of translating the ambient (for example, those aspects seen in Fig.\ref{fig:figure1}) information that metaversae environments show so easily to normative users. At the time, these solutions were hampered by technological gaps, such as a lack of suitable language transformers, and `Moore's Law' governing CPU clock speed increases required by CPU intensive designs.

However, we believe nonetheless that an opportunity for a storytelling device, perhaps using a large language model (such as a generative language model like GPT3) alongside visualization (such as \cite{ohene2016drawing}) and other techniques (for example, novel research findings such as knowledge about room size efficacy potential \cite{shin2019any}), these that have proven their efficacy in other AAT adaptations could be combined to facilitate `storytelling', which in our case means translating the ambient elements of a metaverse into something that facilitates access to disabled individuals. This could serve to enhance the self-efficacy of their interactions with said metaverse. In the physical world, museums often take such combined storytelling tactics in an effort to increase accessibility, and are turning their attention to the metaverse \cite{hutson2023museums}.  

Regardless of the efficacy potential of a solution such as this one, any potential work will need to strive towards shared standards across metaversae in these assistive technologies. Researchers such as Holloway have looked extensively into the issues faced by disabled individuals in terms of technological access (including beyond the metaverse), and have found that due to the heterogeneity of issues facing this group, developing technological solutions tends to be difficult \cite{holloway2019disability}. However, as Homewood et all point out, solving technological problems with yet more technological solutions tends to lead to “missionary” system designers, elbowing their way into situations with the explicit aim of immediately correcting perceived injustices with more of the same \cite{homewood2021tracing}. But there is cause for optimism. The HCI space is in a privileged position, we are a young paradigm, with formative work such as Picard's book \textit{Affective Computing} \cite{picard2000affective} being less than 30 years old at time of writing. We have the opportunity now to make use of these lessons, won so hard in other fields, and utilise them from the beginning. We therefore view these areas not as insurmountable, but as challenges. 

\section{Conclusion}

It typically takes decades for marginalized elements of society to gain access to novel technologies. For example, despite the first long-distance car journey being undertaken by a woman, it more than a century before women were taken into account during crash safety testing \cite{crashtest}. Similarly, in architecture, it took the publishing of Goldsmith's 1963 book \textit{Designing for the disabled} to create a paradigm shift in the way buildings saw inclusivity. In a 2012 annex, Goldsmith reported on these changes, and strove to move to the next phase, where `designing for everyone' would become the next shift \cite{goldsmith2012designing}.

What we hope this paper has shown is that the issues facing marginalized groups in metaversae are derived not simply from technology restrictions, but can trace their origins much further back, and are interdisciplinary in nature. At the same time, the concepts discussed are not inherently exclusionary, and can allow us a better understanding of where these issues come from. By creating conceptual, metaphorical and physical connections between individuals with disabilities, academics, designers, authors, and policymakers, we have the opportunity to surpass the current working paradigm in assistive HCI. It is our hope therefore that this paper may serve as a starting point for broader conversations on how individuals with disabilities can be empowered to create and self-actualize their own virtuality without limitation.

\section{Acknowledgements}

This project is part of AT2030, a programme funded by UK Aid and led by the Global Disability Innovation Hub. AT2030 will test ‘what works’ to improve access to AT and will invest £20m to support solutions to scale. With a focus on innovative products, new service models, and global capacity support, the programme will reach 9 million people directly and 6 million more indirectly to enable a lifetime of potential through life-changing assistive technology. More information at AT2030.org. 

\balance{}

% BALANCE COLUMNS
\balance{}

% REFERENCES FORMAT
% References must be the same font size as other body text.
\bibliographystyle{SIGCHI-Reference-Format}
\bibliography{references.bib}

\end{document}